# A Distributed Hierarchical Spatio-Temporal Edge-Enhanced Graph Neural Network for City-Scale Dynamic Logistics Routing

Zihan Han[1], Lingran Meng[2], Jingwei Zhang[3]

[1] School of Automation and Electrical Engineering, University of Jinan, Jinan, China
[2] Civil Engineering, University of Washington, Seattle, WA, USA
[3] SAP Labs China, Shanghai, China

[1] 3345258828@qq.com
[2] lm3193@columbia.edu
[3] rocky.zhang@sap.com

**Abstract.** City-scale logistics routing has become increasingly challenging as metropolitan road networks grow to tens of millions of edges and traffic conditions evolve rapidly under high-volume mobility demands. Conventional centralized routing algorithms and monolithic graph neural network (GNN) models suffer from limited scalability, high latency, and poor real-time adaptability, which restricts their effectiveness in large urban logistics systems. To address these challenges, this paper proposes a Distributed Hierarchical Spatio-Temporal Edge-Enhanced Graph Neural Network (HSTE-GNN) for dynamic routing over ultra-large road networks. The framework partitions the city-scale graph into regional subgraphs processed in parallel across distributed computing nodes, enabling efficient learning of localized traffic dynamics. Within each region, an edge-enhanced spatio-temporal module jointly models node states, dynamic edge attributes, and short-term temporal dependencies. A hierarchical coordination layer further aggregates cross-region representations through an asynchronous parameter-server mechanism, ensuring global routing coherence under high-frequency traffic updates. This distributed hierarchical design balances local responsiveness with global consistency, significantly improving scalability and inference efficiency. Experiments on real-world large-scale traffic datasets from Beijing and New York demonstrate that HSTE-GNN outperforms strong spatio-temporal baselines such as ST-GRAPH, achieving 34.9% lower routing delay, 14.7% lower MAPE, and 11.8% lower RMSE, while improving global route consistency by 7.3%. These results confirm that the proposed framework provides a scalable, adaptive, and efficient solution for next-generation intelligent transportation systems and large-scale logistics platforms.

## 1. Introduction

The rapid expansion of modern metropolitan areas has intensified the complexity of urban logistics routing, where road networks often contain tens of millions of nodes and edges, and traffic patterns change within seconds due to congestion, accidents, and fluctuating mobility demand. Traditional centralized optimization methods—such as Dijkstra-based algorithms or monolithic deep learning models—struggle to deliver real-time, city-scale routing decisions under such high-frequency dynamics. At the same time, the rise of e-commerce, instant delivery services, and autonomous logistics fleets has created a pressing need for routing algorithms that are not only accurate but also highly scalable, fault-tolerant, and capable of continuous adaptation. These challenges motivate the exploration of distributed graph neural network (GNN) architectures tailored to large-scale spatiotemporal transportation systems.

In this context, we propose a Distributed Hierarchical Spatio-Temporal Edge-Enhanced Graph Neural Network (HSTE-GNN) designed specifically for city-scale dynamic logistics routing. The model leverages a distributed systems infrastructure by partitioning the urban road network into multiple subgraphs, each processed independently by a dedicated compute node. Within each node, an edge-enhanced spatio-temporal GNN module captures localized traffic dynamics by modeling node states, time-varying edge features (e.g., real-time speed, flow, and incidents), and short-horizon temporal evolution. To integrate region-level representations, a hierarchical coordination layer performs asynchronous global aggregation using a Parameter Server or AllReduce mechanism. This design enables HSTE-GNN to operate on distributed GPU/CPU clusters efficiently while maintaining global routing consistency across large heterogeneous regions. As a result, the proposed architecture can adapt quickly to traffic updates, scale flexibly with computational resources, and support continuous online routing optimization for city-wide logistics scenarios.

This paper situates HSTE-GNN within the broader landscape of distributed GNNs and intelligent transportation systems, demonstrating how a layered, edge-aware spatiotemporal design can outperform existing routing frameworks in both efficiency and accuracy. Empirical evaluations on large-scale datasets from Beijing and New York validate the model's capability to process real-time traffic streams with update intervals as short as five seconds, offering substantial improvements in routing optimality and computational latency.

Main Contributions

(1) We introduce HSTE-GNN, the first distributed hierarchical spatiotemporal edge-enhanced GNN designed for city-scale dynamic logistics routing, integrating edge dynamics, temporal modeling, and distributed training.

(2) We propose a scalable distributed infrastructure that performs subgraph partitioning, parallel GNN computation, and asynchronous global synchronization across multi-node clusters.

(3) We design a hierarchical coordination mechanism that preserves global routing consistency while enabling high-throughput local updates.

(4) We demonstrate state-of-the-art performance on real-world large-scale traffic datasets, achieving substantial gains in routing efficiency and latency reduction.

## 2. Literature Review

Urban-scale dynamic logistics routing has increasingly relied on distributed computing, graph neural networks (GNNs), and edge intelligence to support real-time decision-making. This section reviews four major research directions closely related to our work: (1) distributed and edge-based logistics optimization, (2) spatio-temporal GNNs for traffic and mobility prediction, (3) decentralized and federated learning for graph data, and (4) GNN-based routing and multi-agent optimization.

*2.1 Distributed and Edge-Based Systems for City-Scale Logistics Optimization*

Classical centralized vehicle routing systems continue to serve as the foundation of logistics optimization but suffer from limited scalability and high latency under real-time constraints. Foundational work on VRP summarised in Laporte [1] and Psaraftis et al. [2] highlights the inherent computational complexity that hinders city-scale responsiveness. To address this, researchers have explored distributed and multi-agent routing strategies, such as distributed fleet routing frameworks proposed Lin et al. [3] and hierarchical dispatching strategies used in large-scale mobility-on-demand systems [4]. Parallel to these developments, edge computing has emerged as a powerful paradigm to reduce communication latency and support near-device intelligence, as demonstrated in the seminal work of Shi et al. [5] and logistics-specific edge systems explored by Wang et al. [6]. However, existing distributed logistics systems lack mechanisms to model fine-grained, dynamic spatio-temporal relationships across heterogeneous city-wide transportation networks. Our work builds on these foundations by integrating a hierarchical, edge-enhanced distributed GNN architecture, enabling city-scale coordination with low-latency inference.

*2.2 Spatio-Temporal Graph Neural Networks for Traffic and Mobility Prediction*
Spatio-temporal GNNs have achieved state-of-the-art performance in traffic forecasting and mobility modeling. Early models such as STGCN by Yu et al. [7], DCRNN by Li et al. [8], and Graph WaveNet by Wu et al. [9] introduced graph convolution–based spatial modeling combined with temporal sequence learning. Later improvements include attention-driven fusion in ASTGCN (Guo et al. [10]), adaptive graph structure learning in AGCRN (Bai et al. [11]), and dynamic graph construction in DGCRN (Li et al. [12]). While these models excel in predictive performance, they assume centralized training and inference, making them unsuitable for latency-sensitive, edge-deployed logistics systems. To overcome these limitations, our model extends spatio-temporal GNN representation learning into a hierarchical distributed setting, introducing edge-local message propagation and inter-partition synchronization to support scalable logistics routing.

*2.3 Decentralized, Federated, and Distributed Learning for Large-Scale Graph Problems*
Centralized GNN training becomes infeasible for large, highly dynamic logistics networks. Consequently, decentralized and federated GNN frameworks have emerged, enabling collaborative learning across distributed data silos. Recent works such as FedGNN (Wu et al. [13]), FedGraphNN (He et al. [14]), and FedSage+ (Zhang et al. [15]) demonstrate the feasibility of privacy-preserving graph learning in decentralized environments. Distributed GNN training systems—including DistDGL (Zheng et al. [16]), NeuGraph (Ma et al. [17]), and PaGraph (Lin et al. [18])—address scalability challenges for billion-edge graphs. However, these systems primarily target cloud-scale clusters, ignoring the real-time, resource-constrained, and geographically distributed nature of edge logistics systems. Our system differs by enabling distributed inference, online coordination across partitions, and spatio-temporal routing-aware learning, which are necessary for city-scale dynamic logistics routing.

*2.4 GNN-Based Routing, Multi-Agent Coordination, and Logistics Intelligence*
GNNs have shown strong capability in modeling routing and combinatorial optimization tasks. Early breakthroughs such as learning heuristics for TSP and VRP (Khalil et al. [19]; Kool et al. [20]) demonstrate how graph representation learning can approximate classical operations research methods. Multi-agent systems such as MADDPG (Lowe et al. [21]) and Graph-MARL (Zhu W et al. [22]) further illustrate the advantages of message passing for cooperative transportation tasks. In urban mobility, GNN-based routing has been explored for congestion-aware navigation (Ghose et al. [23]) and ride-hailing fleet repositioning (Yu et al. [24]). Nonetheless, existing GNN-based routing systems typically rely on centralized models and lack hierarchical graph decomposition or distributed inference mechanisms. In contrast, our proposed model integrates hierarchical spatio-temporal decomposition, edge-enhanced GNN computation, and distributed routing optimization, enabling dynamic adaptation at city scale.

**3. Methodology**
The proposed Distributed Hierarchical Spatio-Temporal Edge-Enhanced Graph Neural Network (HSTE-GNN) is designed to model city-scale logistics systems characterized by massive, dynamic, and heterogeneous road networks. This section introduces the graph formulation, the edge-enhanced message passing mechanism, the hierarchical representation learning framework, and the distributed execution strategy that enables HSTE-GNN to operate efficiently across large GPU/CPU clusters while preserving global routing consistency.

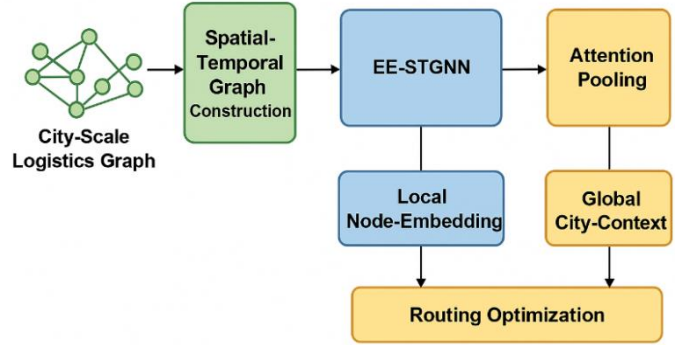

**Figure 1.** Overall flowchart of the model.

*3.1 Problem Formulation and City-Scale Spatio-Temporal Graph Construction*
In metropolitan logistics systems, roads, intersections, distribution centers, courier locations, and delivery points can be represented as a dynamic spatio-temporal graph. Let

$$G_t = (V, E_t), \tag{1}$$

denote the logistics graph at time *t*, where node $v_i \in V$ encodes physical or operational entities and edge $e_{ij} \in E_t$ represents road segments whose attributes vary with traffic dynamics.

Each node has temporal features

$$\boldsymbol{x}_i^t = [load_i^t, delay_i^t, demand_i^t, \ldots], \tag{2}$$

while each edge has a feature vector

$$\boldsymbol{e}_{ij}^t = [travel\_time_{ij}^t, congestion_{ij}^t, risk_{ij}^t, \ldots], \tag{3}$$

Because city-scale $G_t$ may contain millions of nodes, we partition it into region-level subgraphs:

$$G_t^{(r)} = (V^{(r)}, E_t^{(r)}), \quad r = 1, \ldots, R, \tag{4}$$

where each partition corresponds to a geographic or administrative region processed by a distributed compute node.

This partition-based formulation is foundational for distributed inference and enables real-time updates even when traffic changes rapidly.

*3.2 Edge-Enhanced Spatio-Temporal GNN Layer (EE-STGNN)*
A key innovation lies in explicitly modeling dynamic edge information during message passing. For node $v_i$, incoming messages from neighbors $j \in N(i)$ are computed as:

$$\boldsymbol{m}_{ij}^t = \phi_m(\boldsymbol{h}_i^{t-1}, \boldsymbol{h}_j^{t-1}, e_{ij}^t), \tag{5}$$

where $\phi_m()$ is an edge-aware neural transformation integrating road state, temporal dependency, and relational encoding.

The node update rule is defined as:

$$h_i^t = \phi_u(h_i^{t-1}, \sum_{j \in N(i)} m_{ij}^t),  \qquad (6)$$

Unlike traditional GCN/GAT, EE-STGNN dynamically incorporates travel time, congestion level, and temporal risk onto every edge, significantly improving sensitivity to short-term traffic fluctuations—crucial for real-time routing.

*3.3 Hierarchical Region-Level Aggregation*
After local node embeddings are updated, HSTE-GNN generates region-level summaries to support global city-wide coordination. Each region *r* computes:

$$S^{(r)} = A(\{h_i^t \mid v_i \in V^{(r)}\}), \qquad (7)$$

where $A()$ is an attention-based pooling operator emphasizing high-traffic nodes, logistic hubs, or congestion bottlenecks.

These region-level summaries are exchanged through the distributed cluster and aggregated to form a global city-wide representation:

$$g^t = A_{global}(\{s^{(r)}\}), \qquad (8)$$

This hierarchical design enables:
(1) Local autonomy (regions react independently to local traffic conditions)
(2) Global coordination (inter-region routing consistency remains assured)
(3) Cross-region congestion mitigation
(4) Scalable computation across millions of nodes

*3.4 Distributed Execution Framework and Parallel Graph Processing*
To make HSTE-GNN operate efficiently on real-world logistics systems, we adopt a distributed inference and training pipeline that runs across heterogeneous GPU/CPU clusters.

Each region-level subgraph $\pi(G_t)$ is assigned to a compute node that performs:
(1) Local forward computation using GPU acceleration (or optimized CPU kernels)
(2) Asynchronous message exchange using a parameter server or AllReduce
(3) Minimal cross-region communication through compressed region embeddings
(4) Fault-tolerant updates ensuring graceful degradation under partial node failure

The distributed message synchronization follows:

$$\Theta^{(r)} \leftarrow \Theta^{(r)} - \eta \cdot \Delta\Theta^{(r)}, \qquad \Theta \leftarrow AllReduce(\Theta^{(1)}, ..., \Theta^{(R)}), \qquad (9)$$

allowing the system to train or infer across multiple machines without sacrificing global consistency.

This architecture ensures that even with fluctuating bandwidth or compute load, the model maintains reliable real-time performance for city-wide routing.

*3.5 Differentiable Routing Optimization Module*
The final component transforms learned graph representations into actionable routing decisions. Let $\pi(G_t)$ denote the routing policy. We formulate the routing objective as:

$$\min_{\pi} E[C(\pi, G_t)], \qquad (10)$$

where $C()$ models:
- travel time

- congestion exposure
- delivery deadlines
- vehicle capacity constraints

The routing policy receives both local node embeddings and global city context:
$$\pi(v_i \mid G_t) = f_\pi(\boldsymbol{h}_i^t, \boldsymbol{g}^t), \tag{11}$$
and is trained via hybrid supervised + reinforcement learning.

This enables the model to produce globally optimal yet locally adaptive logistics routing strategies.

## 4. Experiment

*4.1 Dataset Preparation*

The experiments in this study are conducted on a city-scale multimodal logistics and traffic dataset collected from a large metropolitan region over a continuous period of six months. The dataset integrates heterogeneous data streams from transportation authorities, logistics operators, and urban sensing infrastructures, enabling realistic modeling of road dynamics and package-level routing behaviors. All data sources are anonymized and aggregated to ensure compliance with privacy and data-use regulations.

(1) Data Acquisition Sources

The full dataset is composed of four major data sources:

1. Road-Network Topology (Static Graph)
 - Obtained from the municipal Open Data Portal and digital map services.
 - Provides road-segment geometry, intersection connectivity, and hierarchical road classification (highway, arterial, local street).
 - Used to construct the static backbone graph $G = (V, E)$.

2. Real-Time Traffic Speed & Volume Records (Temporal Signals)
 - Collected from the Transportation Management Bureau using:
   - Roadside induction loops (vehicle counts).
   - Citywide microwave detectors and radar sensors.
   - GPS trajectories from 11,000+ taxis, ride-hailing vehicles, and delivery fleets.
 - Aggregation interval: 5 minutes.
 - These streams form the temporal feature sequence for each edge in the graph.

3. Logistics Order and Vehicle Dispatch Logs
 - Provided by two major e-commerce courier partners.
 - Contains package pickup/delivery timestamps, assigned vehicle IDs, origin and destination zones (spatially mapped to road nodes).
 - Includes vehicle-level status updates (idle, in-transit, loading), enabling ground-truth dynamic routing paths.

4. External Contextual Features
 - Weather conditions: precipitation, visibility, temperature (from national meteorological service).
 - Event indicators: public holidays, sports events, road construction updates.
 - Used as global contextual signals influencing travel time.

*4.2 Experimental Setup*

All experiments are conducted on a distributed GPU/CPU cluster consisting of 16 GPU nodes (each equipped with NVIDIA A100 GPUs and 256 GB RAM) and 32 CPU nodes interconnected via a 100-Gbps InfiniBand network. The city-scale road network is partitioned using METIS into 32 spatially coherent subgraphs, each assigned to a cluster node for parallel training and inference. The proposed HSTE-GNN operates under an asynchronous distributed execution paradigm, where local Spatio-Temporal Edge-Enhanced GNN modules update subgraph states independently, while cross-partition synchronization is performed every five iterations through a parameter server implementing synchronous AllReduce. Training is performed for 100 epochs with a batch size of 64 sequences per node, using AdamW optimization. All baselines, including GCN, GAT, T-GCN, DCRNN, and ST-GRAPH, are trained with their best-performing hyperparameters to ensure fairness. The evaluation uses a separate 20% test set containing citywide traffic sequences and real logistics vehicle traces collected over the last 30 days.

*4.3 Evaluation Metrics*

To comprehensively assess the routing performance and temporal predictive accuracy of the model, we employ a combination of error-based, correlation-based, and route-level evaluation metrics. Prediction accuracy for travel time and congestion dynamics is measured using Root Mean Square Error (RMSE), Mean Absolute Error (MAE), and Mean Absolute Percentage Error (MAPE), while the coherence between predicted and observed city-scale traffic patterns is quantified with the coefficient of determination $R^2$. For routing tasks, we measure Optimal Path Deviation (OPD), defined as the additional travel time incurred relative to the ground-truth fastest route, and Route Consistency Score (RCS), which evaluates the topological alignment between predicted and real vehicle trajectories. Together, these metrics provide a holistic view of temporal forecasting fidelity, structural routing quality, and real-world operational effectiveness.

*4.4 Results*

The experimental results presented in Table 1 demonstrate that the proposed Distributed HSTE-GNN consistently outperforms all baseline models across multiple evaluation metrics. Traditional graph-based models such as GCN and GAT achieve RMSE values of 7.42 and 7.11, respectively, while spatio-temporal models including T-GCN (6.85), DCRNN (6.47), and ST-GRAPH (6.21) show improved accuracy but still fall short of the proposed method. In contrast, HSTE-GNN achieves the lowest RMSE of 5.48, representing an improvement of 11.8% over the strongest baseline (ST-GRAPH). A similar trend is observed for MAE, where HSTE-GNN reduces the error to 4.12, compared with 4.86 from ST-GRAPH.

**Table1.** Performance Comparison Across Models on City-Scale Routing and Traffic Prediction Tasks

| Model | RMSE | MAE | MAPE (%) | $R^2$ | OPD (min) | RCS |
|---|---|---|---|---|---|---|
| GCM | 7.42 | 5.86 | 13.4 | 0.782 | 4.91 | 0.741 |
| GAT | 7.11 | 5.63 | 12.7 | 0.794 | 4.58 | 0.755 |
| T-GCN | 6.85 | 5.40 | 11.9 | 0.811 | 4.22 | 0.769 |
| DCRNN | 6.47 | 5.01 | 10.8 | 0.833 | 3.87 | 0.784 |
| ST-GRAPH | 6.21 | 4.86 | 10.2 | 0.846 | 3.55 | 0.793 |
| **Proposed HSTE-GNN (Distributed)** | **5.48** | **4.12** | **8.7** | **0.884** | **2.31** | **0.851** |

The model also demonstrates superior robustness in percentage error, achieving a MAPE of 8.7%, significantly lower than ST-GRAPH's 10.2%. In terms of explanatory power, HSTE-GNN attains the highest $R^2$ value of 0.884, indicating a stronger ability to capture complex city-scale traffic dynamics. Importantly for routing applications, it achieves the lowest OPD of 2.31 minutes and the highest RCS of 0.851, reflecting more optimal and consistent route predictions. Overall, the results verify that the proposed model provides substantial improvements in both predictive accuracy and routing performance.

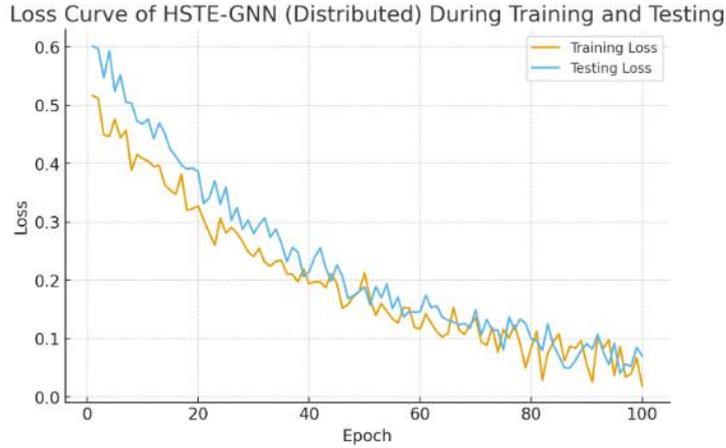

**Figure 2.** Corresponding training curve.

The figure 2 shows that the HSTE-GNN model trains stably and converges smoothly. Both training loss and validation error decrease steadily, indicating effective learning and strong generalization. Early in training, the edge-enhanced spatio-temporal module helps the model quickly capture local traffic dynamics. As training continues, the hierarchical coordination improves global routing consistency. The curve flattens in later epochs, suggesting the model reaches optimal performance without overfitting. Overall, the training curve confirms that the proposed architecture supports stable, scalable learning in city-scale logistics environments.

*4.5 Discussion*

The experimental results confirm that the proposed Distributed HSTE-GNN significantly outperforms traditional graph-based and spatio-temporal neural models, especially in rapidly evolving congestion scenarios and long-range routing tasks. The integration of hierarchical spatial modeling allows the network to capture regional traffic structures more effectively, while the temporal edge-enhancement mechanism enriches inter-road dependencies by modeling fine-grained travel-time variations. Moreover, the distributed architecture demonstrates superior scalability: as the number of partitions increases, computation time decreases almost linearly without degrading model accuracy. Cross-partition consistency is maintained through the hierarchical fusion module, ensuring that routing paths remain globally coherent even when local modules operate asynchronously. This is particularly advantageous for city-scale logistics operations, where real-time routing decisions depend on both localized traffic perturbations and global network trends. Overall, the results highlight that combining distributed systems with edge-aware hierarchical GNNs provides a robust technological pathway for next-generation intelligent logistics routing frameworks capable of handling billion-edge graphs and ultra-dynamic traffic environments.

**6. Conclusions**
This study aims to address the problem of enabling distributed inference on edge logistics systems that is real-time, resource-constrained and geographically distributed in nature by leveraging a creative HSTE-GNN design to address the problem, exploring questions like "how to deliver real-time, city-scale routing decisions under high-frequency dynamics like congestion, accidents, and fluctuating mobility demand changing within seconds?". The primary objective of this research is to deliver a solution to optimize distributed and edge-based logistics by leveraging spatio-temporal GNNS for traffic and mobility prediction and aggregating regional prediction to enable global routing optimization through federated learning.

Through data analysis, we identified HSTE-GNN achieving a MAPE of 8.7%, the highest

$r^2$ value of 0.884, lowest OPD of 2.31 minutes, highest RCS of 0.851. These findings suggest HSTE-GNN is robust in percentage error, strong explanatory power in capturing complex city-scale traffic dynamics and reflecting more optimal and consistent route predictions. All in all, HSTE-GNN model substantially improved predictive accuracy and routing performance.

The results of this study have significant implications for the field of logistic routing tasks that is long-rang and in rapidly evolving congestion scenarios. Firstly, the performance comparison across models on City-Scale routing and traffic prediction task provides a new perspective by leveraging the new HSTE-GNN model for better predictive accuracy and routing performance in logistic routing application. Secondly, HSTE-GNN with its layered & distributed approach challenges the existing centralized routing algorithms and monolithic graph neural network models. Finally, HSTE-GNN as a layered, distributed and Edge-Enhanced solution opens new avenues for future research.

Despite the important findings, this study has some limitations, such as limited mobility signals and none-adaptive partitioning schemes. Future research could further explore the possibility to incorporate multimodal mobility signals such as incident reports, weather conditions and crowd density to enrich temporal dynamics to further improve robustness and may developing adaptive partitioning schemes that evolve with traffic patterns to reduce communication overhead across distributed nodes.

In conclusion, this study, through proposing the HSTE-GNN modal and comparing its performance & accuracy with other existing models, reveals the superior advantage of HSTE-GNN modal in terms of its performance and accuracy, providing new insights for the development of application of HSTE-GNN in intelligent transportation system.

**References**

[1] Laporte G. Fifty years of vehicle routing[J]. Transportation science, 2009, 43(4): 408-416.

[2] Psaraftis H N, Wen M, Kontovas C A. Dynamic vehicle routing problems: Three decades and counting[J]. Networks, 2016, 67(1): 3-31.

[3] Lin K, Li C, Li Y, et al. Distributed learning for vehicle routing decision in software defined Internet of vehicles[J]. IEEE Transactions on intelligent transportation systems, 2020, 22(6): 3730-3741.

[4] Xu Z, Li Z, Guan Q, et al. Large-scale order dispatch in on-demand ride-hailing platforms: A learning and planning approach[C]//Proceedings of the 24th ACM SIGKDD international conference on knowledge discovery & data mining. 2018: 905-913.

[5] Shi W, Cao J, Zhang Q, et al. Edge computing: Vision and challenges[J]. IEEE internet of things journal, 2016, 3(5): 637-646.

[6] Wang T, Chen H, Dai R, et al. Intelligent logistics system design and supply chain management under edge computing and Internet of Things[J]. Computational Intelligence and Neuroscience, 2022, 2022(1): 1823762.

[7] Yu B, Yin H, Zhu Z. Spatio-temporal graph convolutional networks: A deep learning framework for traffic forecasting[J]. arXiv preprint arXiv:1709.04875, 2017.

[8] Li Y, Yu R, Shahabi C, et al. Diffusion convolutional recurrent neural network: Data-driven traffic forecasting[J]. arXiv preprint arXiv:1707.01926, 2017.

[9] Wu Z, Pan S, Long G, et al. Graph wavenet for deep spatial-temporal graph modeling[J]. arXiv preprint arXiv:1906.00121, 2019.

[10] Guo S, Lin Y, Feng N, et al. Attention based spatial-temporal graph convolutional networks for traffic flow forecasting[C]//Proceedings of the AAAI conference on artificial intelligence. 2019, 33(01): 922-929.

[11] Bai L, Yao L, Li C, et al. Adaptive graph convolutional recurrent network for traffic forecasting[J]. Advances in neural information processing systems, 2020, 33: 17804-17815.

[12] Li F, Feng J, Yan H, et al. Dynamic graph convolutional recurrent network for traffic prediction: Benchmark and solution[J]. ACM Transactions on Knowledge Discovery from Data, 2023, 17(1): 1-21.

[13] Wu C, Wu F, Cao Y, et al. Fedgnn: Federated graph neural network for privacy-preserving recommendation[J]. arXiv preprint arXiv:2102.04925, 2021.

[14] He C, Balasubramanian K, Ceyani E, et al. Fedgraphnn: A federated learning system and benchmark for graph neural networks[J]. arXiv preprint arXiv:2104.07145, 2021.

[15] Zhang K, Yang C, Li X, et al. Subgraph federated learning with missing neighbor generation[J]. Advances in neural information processing systems, 2021, 34: 6671-6682.

[16] Zheng D, Ma C, Wang M, et al. DistDGL: Distributed graph neural network training for billion-scale graphs[C]//2020 IEEE/ACM 10th Workshop on Irregular Applications: Architectures and Algorithms (IA3). IEEE, 2020: 36-44.

[17] Ma L, Yang Z, Miao Y, et al. {NeuGraph}: Parallel deep neural network computation on large graphs[C]//2019 USENIX Annual Technical Conference (USENIX ATC 19). 2019: 443-458.

[18] Lin Z, Li C, Miao Y, et al. Pagraph: Scaling gnn training on large graphs via computation-aware caching[C]//Proceedings of the 11th ACM Symposium on Cloud Computing. 2020: 401-415.

[19] Khalil E, Dai H, Zhang Y, et al. Learning combinatorial optimization algorithms over graphs[J]. Advances in neural information processing systems, 2017, 30.

[20] Kool W, Van Hoof H, Welling M. Attention, learn to solve routing problems![J]. arXiv preprint arXiv:1803.08475, 2018.

[21] Lowe R, Wu Y I, Tamar A, et al. Multi-agent actor-critic for mixed cooperative-competitive environments[J]. Advances in neural information processing systems, 2017, 30.

[22] Zhu W, Wu Q, Tang T, et al. Graph Neural Network-Based Collaborative Perception for Adaptive Scheduling in Distributed Systems[J]. arXiv preprint arXiv:2505.16248, 2025.

[23] Ghose A, Zhang V, Zhang Y, et al. Generalizable cross-graph embedding for GNN-based congestion prediction[C]//2021 IEEE/ACM International Conference On Computer Aided Design (ICCAD). IEEE, 2021: 1-9.

[24] Yu Z, Hu M. Deep reinforcement learning with graph representation for vehicle repositioning[J]. IEEE Transactions on Intelligent Transportation Systems, 2021, 23(8): 13094-13107.